\pdfoutput=1
\documentclass[aps,pra,amsmath,amssymb,amsfonts,reprint,groupedaddress,showpacs,showkeys,floatfix]{revtex4-1}

\usepackage[utf8]{inputenc}
\usepackage{siunitx}
\usepackage{physics}
\usepackage{xfrac}
\usepackage{graphicx}
\usepackage{revsymb4-1}
\usepackage{bm}
\usepackage{times}
\usepackage{soul}
\usepackage{ulem}

\usepackage[usenames,dvipsnames]{xcolor}

\usepackage{layouts} 

\newcommand{\yb}[1]{\textsuperscript{#1}Yb}
\newcommand{\ybion}[1]{\textsuperscript{#1}Yb\textsuperscript{+}}

\renewcommand{\vec}[1]{{\mathbf{#1}}}

\newcommand{\fidelity}[1][]{{\mathcal{F}}_{#1}}
\newcommand{\state}[3]{{#1}_{\sfrac{#2}{#3}}}

\begin{document}

\normalem


\title{High-Fidelity Preservation of Quantum Information During Trapped-Ion Transport}
\author{Peter~Kaufmann}
\author{Timm~F.~Gloger}
\author{Delia~Kaufmann}
\author{Michael~Johanning}
\author{Christof~Wunderlich}

\email{christof.wunderlich@uni-siegen.de}
\homepage{http://quantenoptik.uni-siegen.de}
\affiliation{Department Physik, Naturwissenschaftlich-Technische Fakultät, Universität Siegen, 57068 Siegen, Germany}

\date{\today}

\begin{abstract}
A promising scheme for building scalable quantum simulators and computers is
the synthesis of a scalable system using interconnected subsystems.
A prerequisite for this
approach is the ability to faithfully transfer quantum information between subsystems. With
trapped atomic ions, this can be realized by transporting ions with quantum  information
encoded into their internal states. Here, we measure with high precision the fidelity of quantum information encoded into hyperfine
states of a \ybion{171} ion during ion transport in a microstructured Paul
trap. Ramsey spectroscopy of the ion's internal state is interleaved  with up to
$4000$ transport operations over a distance of \SI{280}{\micro m} each taking
\SI{12.8}{\micro s}. We obtain a state fidelity of $99.9994\left({}^{+6}_{-7}\right)$\%
per ion transport.

\end{abstract}


\maketitle
Ion traps have been a workhorse in demonstrating many proof-of-principle experiments in quantum information processing using small ion
samples \cite{Blatt2008}. A major challenge to transform this ansatz into a
powerful quantum computing machine that can handle problems beyond the capabilities of
classical super computers remains its scalability \cite{Kielpinski2002,Svore2006,Monroe2014,Lekitsche2017}. Error correction
schemes allow us to fight the ever sooner death of fragile quantum
information stored in larger and larger quantum systems, but their
economic implementation requires computational building blocks to be
executed with sufficient fidelity~\cite{Shor1995,Steane1996}.
Essential  computational steps
have been demonstrated with fidelities beyond a threshold of $99.99 \%$ that is often considered as allowing for economic error correction \cite{Knill2010}, and, thus for fault-tolerant scalable quantum information processing (QIP). These building blocks include  
single qubit rotation \cite{Brown2011,Ballance2011},
 individual addressing of interacting ions ~\cite{Piltz2014}, and internal state detection~\cite{Burrell2010}.
In addition, high fidelity two-qubit
quantum gates~\cite{Benhelm2008,Ballance2011,Harty2016,Gaebler2016,Weidt2016} and coherent three-qubit conditional quantum gates \cite{Monz2009,Piltz2016} have been implemented.

Straightforward scaling up to an arbitrary size of a single ion trap quantum register, at present, appears unlikely to be successful because the growing size
of a single register usually introduces additional constraints imposed by the confining potential and
by the Coulomb interaction of ion strings \cite{Johanning2016}. Even though, for instance,  transverse modes and anharmonic
trapping~\cite{Lin2009} may be employed for conditional quantum logic, a general claim might be that,
at some point it is useful to divide a single ion register into subsystems and to exchange
quantum information between these subsystems \cite{Kielpinski2002,Svore2006,Monroe2014,Lekitsche2017}.
One might do that by transferring quantum information from ions to photons (and vice versa) and by then exchanging photons between subsystems \cite{Monroe2014,Hucul2015}.

Alternatively, when exchanging quantum information between spatially separated individual registers
within an ion trap-based quantum information processor, the transport of ions carrying this information is an attractive approach \cite{Kielpinski2002,Svore2006,Lekitsche2017}.
Methods to transport ions in segmented Paul traps have been developed and demonstrated \cite{Rowe2002,Hensinger2006,Blakestad2009,Singer2010},
and optimized with respect to the preservation of the motional state during
transport~\cite{Bowler2012,Walther2012}.

It is equally important to avoid errors of the quantum information
encoded into internal states of ions during transport.
Schemes relying on physical transport of ions require shuttling of ions between
regions where the actual conditional gates take place (or between memory zones).
Transport and single qubit manipulation can also be combined and
executed at the same time~\cite{DECLERCQ2016}.

Quantum error correction relies on the distribution of a logical qubit's information
onto multiple qubits. Encoding and correction of this information consists,
in general, of a number of single qubit rotations, entangling gates, measurements,
and typically either shuttling or spectroscopic decoupling of ions. To have
the entire error correction sequence be beneficial, the constraints on the individual
operations are obviously more stringent.
For all correction schemes involving ion transport, the number of transport operations
are bigger compared to or much larger than one \cite{Kielpinski2002,Chiaverini2005,Svore2006,Lekitsche2017}, so
the infidelity must, at least, be an order of magnitude
smaller than acceptable for the entire sequence.

Therefore, in addition to high fidelity local gates, high fidelity transport is
required to not cross a desired error threshold when carrying out single- and
multiqubit quantum gates.

Several experiments have characterized
the internal state fidelity $\fidelity = \expval{\rho}{\psi}$ upon transport
by measuring the loss of coherence of a prepared superposition state $\ket{\psi}$ which dephases into a
mixed state $\rho$ during a Ramsey-type measurement. However, the precision reached in these experiments was not yet sufficient to conclude that transport takes place in the fault-tolerant regime required for scaling
\cite{Rowe2002,Blakestad2009,Bowler2012,Walther2012}\footnote{We assume
gaussian errors of the reported Ramsey fringe contrasts and calculate the fidelity as
$\fidelity = \sqrt[M]{C_M/C_0})$, where $C_0$ is the contrast before and $C_M$ after
$M$ transport operations.}.
Here, we demonstrate high fidelity
transport of trapped ions over a distance of \SI{280}{\micro m} with quantum information encoded into internal hyperfine states with a relative error of the qubit states per transport below $10^{-5}$ which is compatible with fault-tolerant and, thus, scalable quantum computation.

The determination of $\fidelity$ is limited by the uncertainty of the extracted Ramsey-fringe
contrast and the relative error is of the same order as the relative uncertainty
of the contrast. In the experiments reported below, we determine the contrast of a Ramsey-type
measurement
typically with a relative error
of \num{\leq 1e-2}.
Therefore, the straightforward extraction of fringe contrast from experimental
data is not sufficient for precise determination of the error taking place
during transport. To be able to precisely measure the loss of fidelity, we
increase the number of transport operations $M$. To limit systematic errors due
to a possible spatial variation of the qubit coherence time in the trap, we design
the experiment such that the ions' average position is independent of $M$ for
$M>0$ and compare the contrast after $M=4000$ with the contrast obtained after
$M=2$ transport operations.

\begin{figure}
  \includegraphics[]{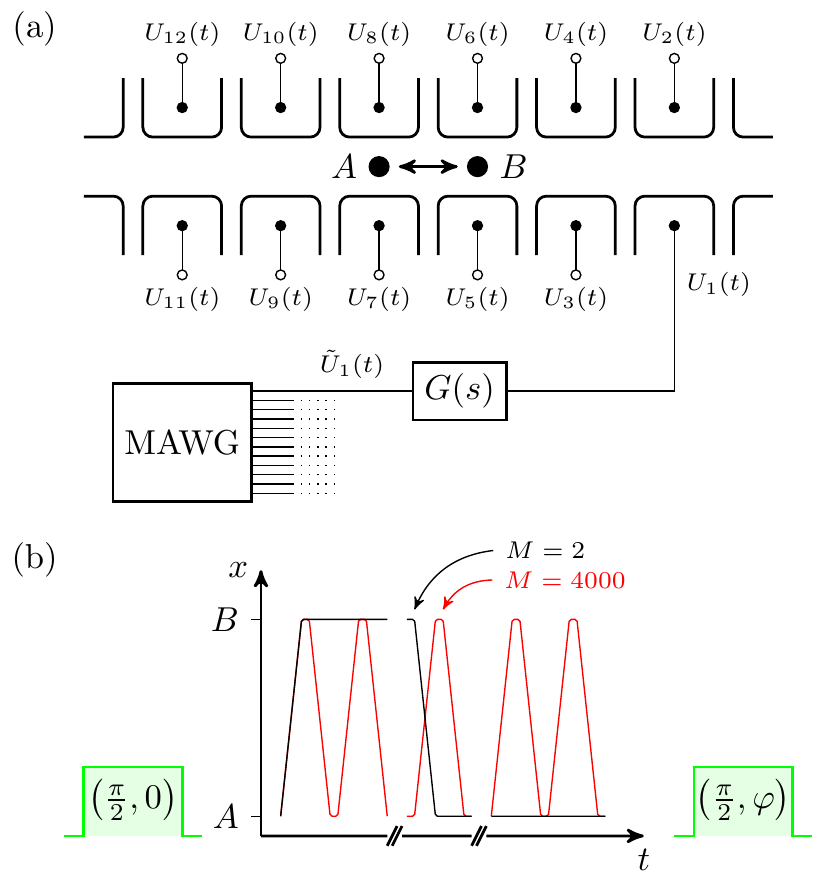}
\caption{Schematic of the experiment. {\it(a)} The voltages $\tilde{U}(t)$ at the multichannel arbitrary waveform generator (MAWG) take the filter characteristic $G(s)$ of the electronics into account
to produce suitable voltages $U(t)$ that transports the ion along the x axis $M$ times
between positions $A$ and $B$. {\it(b)} The timing of the transport operations is such that the
ion is located equally long at $A$ and $B$. The ion transfer is sandwiched by two
$\pi/2$-pulses.}
\label{fig:setup}
\end{figure}

The ion trap is operated with singly charged \yb{172} and \yb{171} ions.
\ybion{172} ions possessing no hyperfine structure are employed to determine the efficiency of physical ion transport, while  for the analysis of the transport induced
decoherence, a hyperfine qubit  in \ybion{171} is used.
As a qubit, we choose
the first-order magnetic field insensitive hyperfine qubit composed of the
states $\ket{0}\equiv\ket{S_{\sfrac{1}{2}},F=0}$ and
$\ket{1}\equiv\ket{S_{\sfrac{1}{2}},F=1,m_F=0}$.
The second order magnetic field sensitivity of the qubit resonance frequency at
$B=\SI{640}{\micro\tesla}$ is $d\nu/d B = \SI{39.7}{\mega\Hz\per\tesla}$.
A detailed description of the laser and microwave setup can be found in
\cite{Vitanov2015}.

The experiment is carried out in a 3D-Paul trap
\cite{Kaufmann2012,Schulz2006,Schulz2008,Baig2013}
divided into 33 segments,
each consisting of two dc electrodes, and two global rf electrodes.
In the experiments reported here, ions were transported by moving the minimum of the trapping potential from the center of one segment $A$
to the center of the next segment $B$ over a distance of \SI{280}{\micro m} (see Fig.~\ref{fig:setup}). The
required potentials were generated by applying twelve voltage ramps to dc electrodes
of six trap segments of the trap.
One additional voltage ramp was applied to a correction electrode
to allow for minimization of micromotion perpendicular to the dc electrodes' plane.

Methods to transport ions in segmented Paul traps have been investigated
elaborately in a number of publications \cite{Rowe2002,Hensinger2006,Blakestad2009,Blakestad2011,Walther2012}.
We use the approach worked out in Ref. \cite{Torrontegui2011} to calculate an optimized trajectory
$\vec{r}(t)$ that minimizes ion heating during transport, an implementation of the
boundary element method to simulate the potentials generated by our trap geometry \cite{Singer2010}, and the formalism
described in Ref.~\cite{Blakestad2011} to calculate suitable transport voltages.
To reduce limitations of the potential dynamics imposed by low pass
filters of our dc electrodes, we extend
this formalism by a method to constructively take into account the filter characteristics:
Instead of trying to compensate the filter behavior after the transport
voltage ramps $U(t)$ have been determined, we calculate the
accessible voltage range for every point $\vec{r}_i = \vec{r}(t_i)$ during transport based on the voltage history $U(t_{k<i})$ and limit the potential
optimization algorithm to this interval. Using this approach, we are able
to realize single transport times of \SI{12.8}{\micro\second} on the order of
the inverse filter cut off frequency (\SI{15.8}{\micro\second}).
See Supplemental Material (Section~\ref{appendix:transport} on page~\pageref{appendix:transport}) for details.

In the experiment presented in this Letter, we performed \num{22e6}
transport operations without losing an ion.
The success of a single transport operation $A\rightarrow B$ or $B\rightarrow A$ is proven by
imaging the ion fluorescence for a few milliseconds once the ion is at rest after transport.
Because up to $2000$
consecutive transport operations $A\rightarrow B \rightarrow A$ are investigated, the success of the overall transport (i.e., during $M$ shuttling events)
needs to be shown as well. Imaging the ion for several milliseconds at one position after
every second transport operation would add seconds to every single repetition of the experiment, and, more
importantly, change the transport dynamics by doppler-cooling of the ions.
So to diagnose the success rate of consecutive transfers, we implemented an
experiment to track the ion during $M$ shuttling operations.
During the transport operation the electron multiplying charged-coupled device camera takes one single image with an
exposure time equal to the overall transport duration. Synchronized to the ion
transport, we flash the detection laser at position $A$ ($B$) for \SI{2}{\micro s}
each time the ion should be at position $B$ ($A$). The absence of the ion is signified by not detecting scattered resonance fluorescence.
This experiment is not exactly tracking the ion, but it proves that it is not at a
position where it should not be. We also perform measurements that flash the ion at
position $A$ ($B$) when it is expected to be there, but the statistics of these
measurements are a factor $15$ inferior
compared to the more sensitive detection of an absent
ion.
This experiment is carried out using \ybion{172}, to profit from higher fluorescence rates.
The analysis yields that $8\left({}^{+12}_{-8}\right)$ out of $4000$ transport operations
are failing. This number corresponds to a transport fidelity (the probability of
transporting the ion as intended) of \SI{99.8}{\percent}. Combined with the
ions presence after \num{22e6} transport
operations, we interpret the obtained transport fidelity as a lower bound.
See Supplemental Material (Section \ref{appendix:transportFidelity} on page~\pageref{appendix:transportFidelity}) for a detailed analysis
of the transport fidelity.

The central goal of the Letter presented here is to determine the effect of ion transport operations
on the qubit's internal state coherence.
The internal state coherence is determined by performing a Ramsey-type
experiment, where transport operations are executed during the free precession time:
We initialize the qubit of a Doppler-cooled ion at position $A$ in the
$\ket{0}$ state. Using a microwave $\pi/2$ pulse, we prepare
the superposition state $\ket{\psi} = 1/\sqrt{2}(\ket{0}-i\ket{1})$. Next,
the ion is transported $M$ times between the positions $A$ and $B$.
We add waiting times at the positions $A$ and $B$ such that the total precession time $t_p = \SI{69.44}{\ms}$ is
independent of $M$.

The waiting time is equally distributed between
both positions. After a second $\pi/2$ pulse with a phase $\varphi$ relative to the first pulse, the
qubit state is read out.

\begin{figure}
  \includegraphics[]{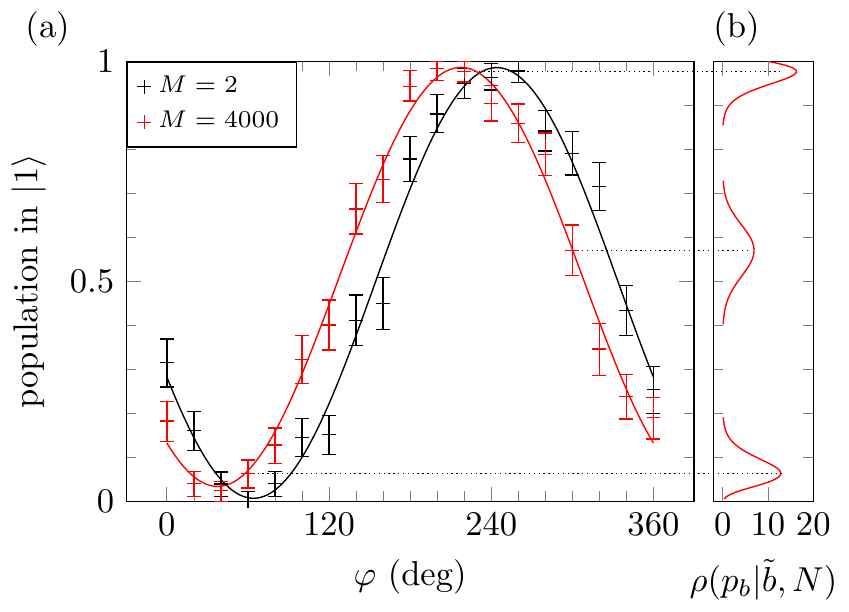}
\caption{{\it(a)} Ramsey fringes for $1$ (black) and $4000$ (red) ion transport
operations. The relative phase $\varphi$ between the two $\pi/2$-pulses is varied while the time between
the pulses is kept constant.
The decay of the amplitude due to the additional transport operations is hardly
noticeable. {\it(b)} Three examples for the probability distributions that were used for the
likelihood analysis of the data.
The probability density $\rho(p_b|\tilde{b},N)$ for $\tilde{b}$ bright events out of $N$ trials is shown
on the horizontal axis as a function of the probability $p_b$ for a projection into
the state $\ket{1}$ on the vertical axis.
}
\label{fig:ramseyFringes}
\end{figure}

We sample the Ramsey interference fringe for $M=2$ and $M=4000$ transport
operations at $19$ phase values $\varphi$. The Ramsey measurement
for every setting is
repeated $100$ times (see Fig.~\ref{fig:ramseyFringes}).

We monitor fluorescence during Doppler-cooling and use a low fluorescence count
(dark or absent ion) as a veto for the last and next Ramsey measurement (about
\SI{10}{\percent} of the data).
Besides electronic readout noise and background counts, the detection scheme is limited by
off resonant excitation of the transitions $\ket{\state{S}{1}{2}, F=0}-\ket{\state{P}{1}{2}, F=1}$ and
$\ket{\state{S}{1}{2}, F=1}-\ket{\state{P}{1}{2}, F=1}$ and following decay into
$\ket{\state{S}{1}{2}, F=1}$ resp. $\ket{\state{S}{1}{2}, F=0}$. The former process results in the observation of
fluorescence from a qubit originally in the $\ket{0}$ (dark) state, the latter one reduces
the fluorescence of the $\ket{1}$ (bright) state. The same effect can be induced by spontaneous decay of
the $\ket{\state{P}{1}{2}, F=0}$ state to the $\ket{\state{D}{3}{2}, F=1}$ state.
By using two separate discrimination thresholds for dark and bright states in the
data analysis, we can reduce the probability of wrongly
identified states at the cost of reduced statistics~\cite{Vitanov2015,Woelk2015}.

For threshold
selection, we add calibration runs to the experiment in which
we omit the $\pi/2$ pulses but prepare
the ion in the $\ket{0}$
($\ket{1}$) state before the transport operations, to obtain detection histograms for pure $\ket{0}$ ($\ket{1}$) states.
The $\ket{1}$ state is prepared by using a BB1RWR $\pi$ pulse \cite{Cummins2003} that is robust against Rabi frequency errors.
The calibration is done separately for both $M=2$ and $M=4000$ transport operations
in order to account for possible variations of the detection statistics due to
transport induced ion heating. Using these calibration measurements, we determine
threshold values for state discrimination. In addition, the probabilities to correctly
identify a bright state as bright $p_{\tilde{b}|b}$ and a dark state as dark $p_{\tilde{d}|d}$
can be extracted. We choose thresholds that yield $p_{\tilde{b}|b} = 0.964$ ($0.959$)
and $p_{\tilde{d}|d} = 0.985$ ($0.978$) for $M=4000$ ($M=2$).

The determination of coherence loss of the qubit state due to
the transport operations is done by comparing the amplitudes of the
obtained Ramsey fringes for different numbers of transport operations $M$. As we expect the infidelity to be
close to zero, we need to employ several statistical methods to get
precise results and error estimates. Since the efficiency of state selective
detection is below unity, we distinguish between the actual number of projections
$b$ and $d$ ($b,d\in\mathbb{N}_0$ and $b+d=N$) into states $\ket{1}$ and $\ket{0}$ and
the corresponding numbers $\tilde{b}$ and $\tilde{d}$ identified as $\ket{1}$ and $\ket{0}$
during data analysis.
To reconstruct the true fractional population of the states $\ket{0}$ and $\ket{1}$
of a qubit state $\ket{\psi}$, we need to infer the numbers  $b$ and $d$ from the
numbers of identified states $\tilde{b}$ and $\tilde{d}$.

The obtained probability density $\rho(p_b|\tilde{b},N)$ of the state
population depends on the number of identified bright states, the number of measurements $N$, and the state
identification probabilities $p_{\tilde{b}|b}$ and $p_{\tilde{d}|d}$.

The state population varies
as a function of the relative phase of the second $\pi/2$ pulse and can be
parameterized by
\begin{equation}
p_b(\varphi) = B + A \sin(\varphi - \Phi)
\label{eq:modelRamsey}
\end{equation}
with the amplitude $A$, offset $B$, and phase shift $\Phi$. We fit this model by maximizing the log likelihood
\begin{equation}
\log \mathcal{L}_M(A,B,\Phi) = \sum_{k=1}^K \log \rho(B + A \sin(\varphi_k - \Phi) | \tilde{b_k}, N_k)
\label{eq:likelihood}
\end{equation}
for both numbers of transport operations $M$ using the probability density function $\rho(p_b| \tilde{b}, N)$
for the $K$ data points $(\tilde{b_k}, N_k)$.

The coherence loss of our qubit in a static potential for
precession times shorter than $\SI{100}{\milli\second}$ is best described by a
decay model $A(t) = \frac{1}{2}\exp\left(-\lambda t^2\right)$ for the amplitude $A$ of the Ramsey fringe
with $\lambda=\SI{4(2)}{\per\square\second}$.
This corresponds to a expected amplitude of $A(t_p)=\num{0.490(5)}$ of a Ramsey
measurement without ion transport.

Figure~\ref{fig:ramseyFringes}(a) shows the Ramsey fringes
obtained for $M=2$ and $M=4000$ transport operations. The error bars
indicate the \SI{68}{\percent} confidence intervals of single data points.
The right part, \ref{fig:ramseyFringes}(b), displays the probability distribution $\rho(p_b|\tilde{b},N)$
for three exemplary data points. The amplitude of the
$M=4000$ curve is slightly reduced by \num{0.013} compared to $M=2$, and the
phase shift differs by \SI{27}{\degree}.
We estimate the gradient of
the magnetic field in the direction of the ion transport to be \SI{10.6e-3}{T/m}.
This gradient results in a \SI{120}{\Hz} difference between the hyperfine splitting of
the qubit at positions $A$ and $B$. From simulations, we expect that the mean positions of the ions for $2$ and $4000$
transfers during the free precession time differ by \SI{0.9}{\mu m}, due to non perfect
compensation of the dc electrode filters. This would correspond to a phase difference
of \SI{10}{\degree}.
The amplitude reduction observed with the number of transport operations is
compatible with zero, in good agreement with our qubit being magnetic field
insensitive to the first order. The Supplemental Material (Section \ref{appendix:decoherence} page~\pageref{appendix:decoherence}) gives
a short discussion of different sources for a possible decay
The state identification probabilities $p_{\tilde{b}|b}$ and $p_{\tilde{d}|d}$
were treated up to this point as fixed error-free parameters. 
In reality these values are calculated from a finite set of measurements and
therefore bear additional uncertainties. We estimate these uncertainties by analyzing
the calibration data obtained for state identification using the bootstrapping
resampling method~\cite{Efron1994} and averaging the likelihoods $\mathcal{L}_M(A)$ over the results obtained for
different choices of $p_{\tilde{b}|b}$ and $p_{\tilde{d}|d}$.
See Supplemental Material (Section \ref{appendix:bootstrap} on page~\pageref{appendix:bootstrap}) for details of the uncertainty estimation
\begin{figure}
  \includegraphics[]{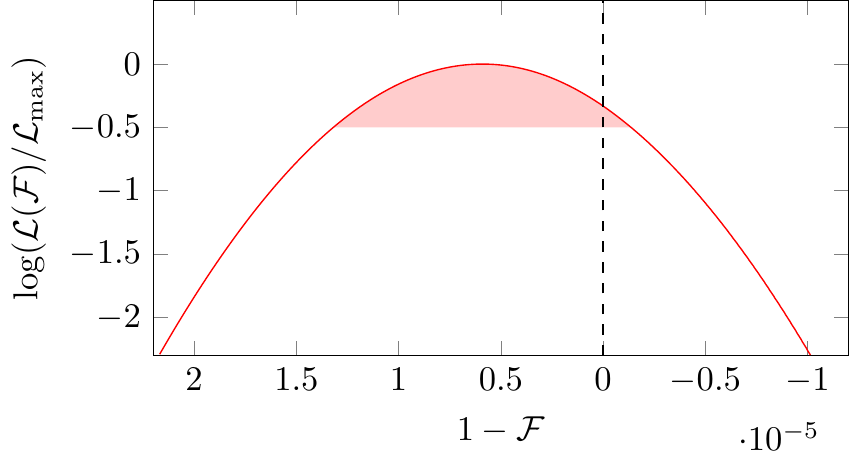}
\caption{Likelihood of the internal state fidelity $\fidelity$ during ion transport.
The loss of internal state fidelity due to the transfer of the ion is apparent but the case of no loss of coherence is also compatible within the $68\%$ confidence
interval (shaded). The log likelihood is calculated by the numerical convolution of the likelihoods $\mathcal{L}_M$.}
\label{fig:internalStateFidelity}
\end{figure}

The likelihood distribution of the internal state fidelity during ion
transport (Fig.~\ref{fig:internalStateFidelity}) is calculated by
a numerical convolution of the likelihoods of the Ramsey fringe amplitudes $A_2$ and $A_{4000}$ according to
\begin{equation}
\fidelity = \sqrt[4000-2]{\frac{A_2}{A_{4000}}}\,.
\end{equation}
Here, we report here a fidelity of the internal qubit state per transport
operation of
\begin{equation}
\fidelity = 0.999994\left({}^{+6}_{-7}\right)\,.
\end{equation}

This result is obtained under the assumption that each individual transport $A\rightarrow B$ and $B\rightarrow A$ out of a total of $M$ attempted transports of an ion is actually successful, that is, the transport fidelity is perfect.
Taking a finite probability of transport failure into account, the fidelity would change according to
$\overline{\fidelity}(M_f) = \fidelity^{(\frac{4000-2}{4000-2-M_f})}$, for $M_f$ failing transports.
The likelihood
of $\fidelity$ is almost gaussian (compare Fig.\ref{fig:internalStateFidelity}), so we
expect the error scaling of $\overline{\fidelity}$ to follow
$\sigma(\overline{\fidelity}(M_f)) = \frac{4000-2}{4000-2-M_f} \fidelity^{(\frac{M_f}{4000-2-M_f})}\sigma(\fidelity)$.
With the transport fidelity of \SI{99.8}{\percent} determined above,
a $5\sigma$ deviation would result in $M_f=68$ failed transports. This would reduce
the internal state fidelity during ion transport by $\num{1e-7}$, which is about
an order of magnitude smaller than the statistical uncertainty of $\fidelity$. A systematic error due to imperfect preparation
of the $\ket{0}$ and $\ket{1}$ states also does not change the statistical significance of $\fidelity$.
See Supplemental Material (Section \ref{appendix:preparationErrors} page~\pageref{appendix:preparationErrors}) for details of the error estimation.

In this work, we use the magnetic insensitive hyperfine qubit. Some schemes for QIP with trapped ions using radio-frequency and microwave radiation \cite{Mintert2001}
utilize magnetic field dependent states, for example, the qubit composed of $\ket{0}$ and
$\ket{S_{\sfrac{1}{2}},F=1,m_F=\pm 1}$. As the magnetic field sensitive qubit can be recoded into the insensitive qubit
and back \cite{Piltz2016} the results of this Letter are also immediately relevant for these QIP schemes.

In summary, we demonstrate  by precise measurements and careful data analysis that the physical transport of quantum information encoded in a hyperfine qubit can be carried out with a fidelity better than $1-10^{-5}$.
This is an important prerequisite, together with high gate fidelities and low cross-talk, for all schemes for scalable QIP with trapped ions that rely on transport of ions.

We acknowledge funding from the European Community's Seventh
Framework Programme under Grant Agreement No. 270843
(iQIT), from EMRP (the EMRP is jointly funded by the EMRP participating countries within EURAMET and the European Union).

\bibliography{bibliography}
\newpage
\appendix{}
\renewcommand{\thesection}{\Roman{section}}

\section{Transport potentials\label{appendix:transport}}
The statistical error of the outcome of our experiment is limited both by the
number of transport operations $M$ and the number of repetitions $N$. As the
transport operations are carried out during the free precession time of a Ramsey
experiment with limited coherence time, $M$ is directly limited by the transfer time.
$N$ is indirectly limited by the overall stability of the experiment apparatus
including microwave and laser power stabilities. In order to improve the
statistics we implement fast ion transport that includes micromotion minimization
and takes electronic filtering into account.

We use the method presented in \cite{Torrontegui2011} with the parameters
axial trap frequency $\omega_x = 2 \pi \times \SI{230}{\kHz}$, transfer distance
$\Delta x = \SI{280}{\micro m}$, a temporal discretization $\Delta t = (\SI{12.5}{\MHz})^{-1} = \SI{80}{\nano s}$ and transport time
$T=160 \cdot \Delta t = \SI{12.8}{\micro s}$ to calculate an ion
trajectory $\vec{r}_i = \vec{r}(t=i\cdot\Delta t)$ discretized in
$L=160$ single steps.

Next a sequence of $L$ potentials is determined with accordingly chosen
potential minima and curvatures at the positions $\vec{r}_i$. For this we first create a set of basis potentials  $\Phi^{(j)}(\vec{r})$ of the single electrodes $j$
which are calculated for all electrodes being grounded except electrode $j$ set to a potential $U_0$
using the boundary element method software package
described in \cite{Singer2010}. The transport trapping potentials $\Phi$ are then obtained as a voltage weighted sum of
the individual basis potentials $\Phi^{(j)}$ and the rf-pseudopotenial $\Phi^{(ps)}$ of
the rf electrodes as presented in \cite{Blakestad2011}.

For every step $i$ the potentials have to fullfill a couple of boundary conditions. In this experiment the position of the
potential minima imposes $3$ requirements ($\partial_{x,y,z}\Phi = 0$), the size of the axial
trap frequency $\partial_x^2 \Phi = {m}\omega_x^2/q$ and the alignment of the
potential axis parallel to the trap's $x$-axis ($\partial_{xz}^2\Phi = 0$) $2$
conditions. We also find it convenient to add a sixth condition
$\Phi(\vec{r}_i) = \Phi_0$ to define the field at the location $\vec{r}_{i}$
of the potential minimum. These
conditions can be expressed as operators $\mathcal{P}$ acting on the potential
$\Phi=U^{(j)} \Phi^{(j)} + \Phi^{(ps)}$ and their corresponding eigenvalues. As we use 12 independent controllable dc electrode potentials the problem is in
principle underconstrained.

In practice the limited voltage range and dynamics of
any real voltage source narrows down the possible solutions and yields
additional constraints. The available minimal and maximal voltages are primary
limiting the accessible potential shapes, while the possible dynamics
limits the potential changing speed. For the experiment reported here the latter
restriction is of primary concern and for simplicity we will not explicitly
write down the constraints by maximal voltages in the following formulas.

If the voltage change per discrete time step is limited
by $\delta U$,
the possible voltage $U_{i}$ for one dc electrode at step $i$
along the trajectory is limited by
\begin{equation}
U_i \lessgtr U_{i-1} \pm \delta U\,.
\label{eq:voltageStepLimit}
\end{equation}

One  can use a constrained optimization algorithm to solve the problem and
obtain the voltages $U_{i}^{(j)}$. If not all boundary conditions can be
fullfilled at the same time in the given voltage limits \eqref{eq:voltageStepLimit} it
can be beneficial to multiply weight factors to the single boundary conditions
and prioritize for example a constant trap frequency over the exact position of the
potential minimum.

\subsection{Modifications for low-pass filter}
It is common practice to low-pass the trap's
dc electrodes in order to reduce the electronic noise in the vincinity of the trapped ions.
If the dynamics of the transport voltage ramps require frequency components near or even above the
cut-off frequency of those filters one needs to
distinguish the voltages $U$ at the trap electrodes from the voltages
$\tilde{U}$ set at the voltage source.

\begin{figure}
  \includegraphics[]{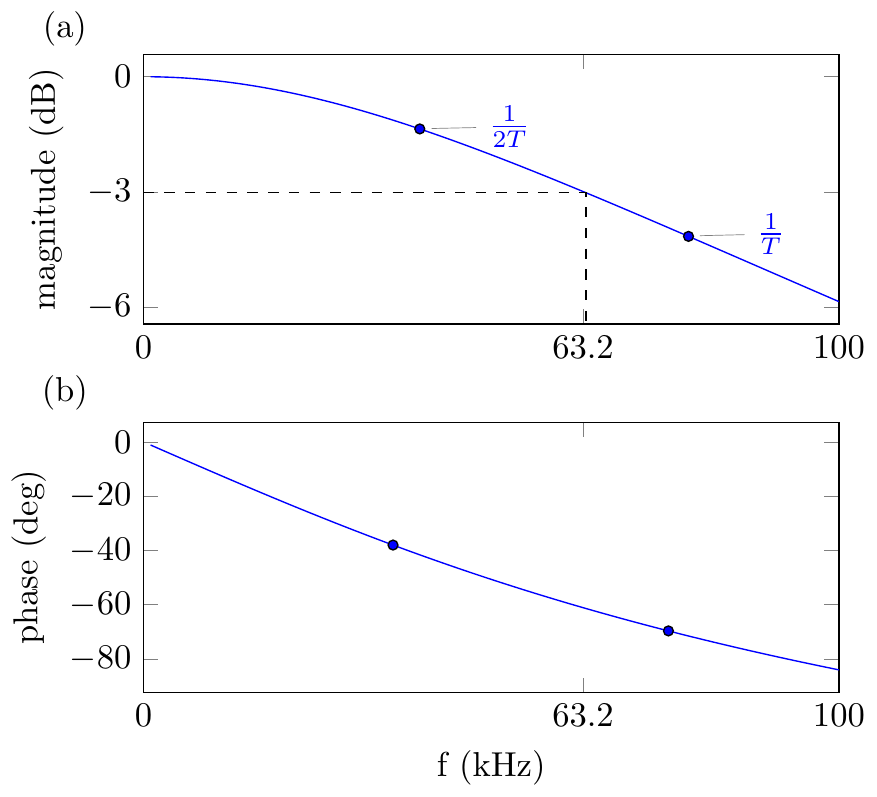}
\caption{Bode plot of the electronic filter for a dc-voltage. The inverse
transport time $1/T$ is on the order of the cut off frequency of the electronic
filters. The horizontal axis of the plot is scaled linear.}
\label{fig:bode}
\end{figure}
One possibility to apply fast step function-like potential changes to trap
electrodes is the usage of switches located next to the trap that are switching
between different voltage channels at a point where the low-pass filtering has
already been applied \cite{Alonso2013}.

An other solution is to calculate backwards source voltages $\tilde{U}(t_i)$ such that one
gets the desired voltages $U(t_i)$ after the low pass filter. A serious downside
of this method is that the maximal possible voltage change $\delta U$
at the electrodes is not only limited by the maximal possible voltage change $\delta\tilde{U}$ of the source, but is also a function of
the history of applied voltages.
To ensure realizable voltage sequences one needs to reduce the value of $\delta U$ in
\eqref{eq:voltageStepLimit} such that in any case it could be produced by a
voltage change of $\delta \tilde{U}$ at the voltage source for all voltage histories.
This precaution --- if realizable at all --- reduces the
available dynamics of the voltage source and in consequence the speed of ion
transport operations.

We circumvent this problem by incorporating the characteristics of the
electronic filters directly in the calculation of the potentials for all steps of
the trajectory.

Our electronic filters are constructed by a series of three
stages of first order RC low-pass filters and the trap capacitance itself: The first stage terminates
our homebuilt programmable voltage source, followed by two stages located next to vacuum interface.
Figure~\ref{fig:bode} shows the corresponding Bode plot. The magnitude of the
amplitude in this plot is determined indirectly by measuring the amplitude of a
sinus signal parallel to the trap and fitting the capacitances of the trap
electrodes. The phase shift is calculated using the obtained model. The cut-off frequency of the system is
\SI{63.2}{\kHz} and on the order of the inverse ion transport time $1/T$, showing
the necessity to take the filter characteristics into account.

From the electronic circuit we obtain the transfer function
$G(s)=\mathcal{L}\{U(t)\}/\mathcal{L}\{\tilde{U}(t)\}$ between source voltage and
electrode voltage as quotient of the two-sided Laplace transforms of input and output voltages with the complex parameter $s=i \omega$ \cite{Oppenheim1989}. From $G(s)$ a time
discrete rational state space transfer function
\begin{equation}
\sum_{n=0}^{n_a} a_n U_{i-n}  = \sum_{n=1}^{n_b} b_n \tilde{U}_{i-n}
\label{eq:filter}
\end{equation}
can be calculated
that connects input and output voltages of the electronic filter  \cite{Oppenheim1989} ($b_0$ is equal zero such that
${U}_{i}$ is independent of $\tilde{U}_{i}$ --- the output is one step behind the
input). $n_a$ and $n_b$ are called
feedback and feedforward filter orders and are equal $4$ in our case. The value of the coefficients $a_n$ and $b_n$ are determined by the resistors and capacitors used.

Solving \eqref{eq:filter} for $U_i$ and substituting $\tilde{U}_{i} = \tilde{U}_{i-1} - \delta \tilde{U}$ resp. $\tilde{U}_{i} = \tilde{U}_{i-1} + \delta \tilde{U}$
yields the range of the
next possible electrode voltages and thus condition \eqref{eq:voltageStepLimit} is replaced by:
\begin{equation}
U_i \lessgtr \frac{1}{a_0} \left( b_1 (\tilde{U}_{i-1} \pm \delta \tilde{U}) + \sum_{n=2}^{n_b} b_n \tilde{U}_{i-n} - \sum_{n=1}^{n_a} a_n U_{i-n} \right)
\label{eq:modifiedVoltageStepLimit}
\end{equation}

\subsection{Micromotion minimization}
The trapping potential simulations assume a perfectly fabricated and
assembled trap and zero electric stray fields. Deviations from these assumptions require
small offset voltages to match the
position of the potentials dc- and rf-null. We determine two sets of optimized offset voltages $\Delta U^{(j)}_1$ and $\Delta U^{(j)}_L$ for
the start and final position of the ion trajectory using the method described by
\cite{Gloger2015}.
For ion positions between these points we are using linearly interpolated values for the offset
voltages $\Delta U_i^{(j)}$. The offset voltages are added to the voltages obtained
from the potential optimization.

In order to guarantee that the total voltage $\overline{U}_i = U_i+\Delta U_i$
including the offset voltages for micromotion minimization
is achievable, the voltage
limits \eqref{eq:modifiedVoltageStepLimit} for the potential problem of the ideal trap
have to be adjusted by $- \Delta U_i$:
\begin{align}
U_i \lessgtr \frac{1}{a_0} &\left( b_1 (\tilde{U}_{i-1} \pm \delta \tilde{U}) + \sum_{n=2}^{n_b} b_n \tilde{U}_{i-n} - \right.\notag\\
 &\left.\sum_{n=1}^{n_a} a_n U_{i-n} - \Delta U_i\right)
\label{eq:modifiedVoltageStepLimitWithCompensation}
\end{align}

The electrostatic problem can now be solved for the electrode voltages $U_i$ and
using \eqref{eq:filter} suitable source voltages $\tilde{\overline{U}}_{i-1}$ including the filter characteristics and micromotion compensation can be calculated:
\begin{equation}
\tilde{\overline{U}}_{i-1} = \frac{1}{b_1} \left( \sum_{n=0}^{n_a} a_n \overline{U}_{i-n} - \sum_{n=2}^{n_b} b_n \tilde{\overline{U}}_{i-n} \right)
\end{equation}

The optimization algorithm can be fine tuned by adding several additional constraints to the
optimization problem. We limit the voltage difference between voltage source and electrodes by
implementing an additional condition $\tilde{\overline{U}} - \overline{U} = 0$ for
the potential optimization problem with a small weight. Additionally we favor
similar voltages for the transport in both directions ($U^{(j)}_{A\rightarrow B}(r_i) - U^{(j)}_{B\rightarrow A}(r_i) = 0$)
to achieve a closed voltage loop, that is needed for an easy scaling of transport operations.
For this condition we use a weight factor that relaxed the condition for coordinates between $A$ and $B$.

\section{Transport fidelity\label{appendix:transportFidelity}}
\begin{figure}
  \includegraphics[]{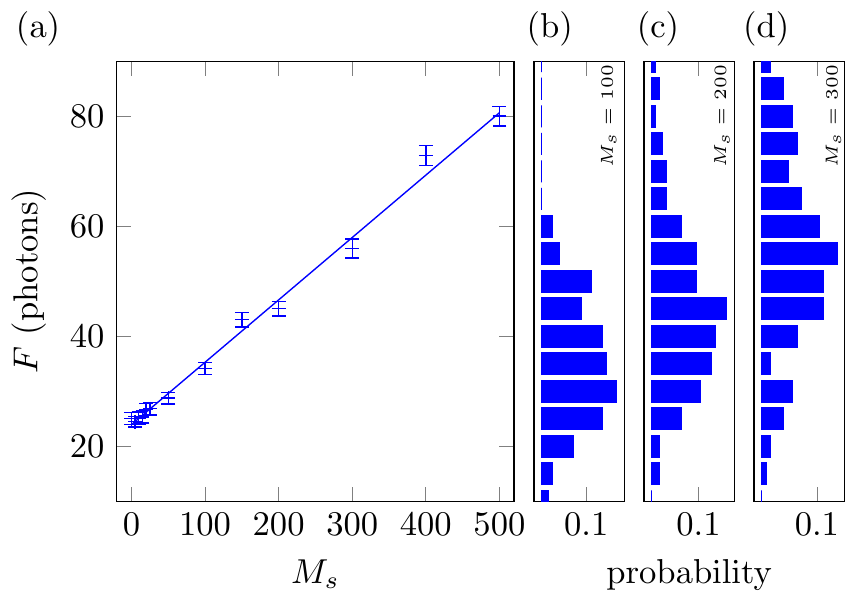}
\caption{{\it (a)} The mean fluorescence $F$ is a linear function of the number $M_s$ of
intentionally skipped transport operations. A linear fit yields \num{0.113(4)} for the slope
and \num{23.9(6)} for the offset of the curve. {\it (b)-(d)} Three sample fluorescence histograms
($M_s=100,200,300$) that were used to extract $F$. Each histogram holds data of 125
single experiment runs.}
\label{fig:shuttleProof}
\end{figure}
We drive the $\ket{S_{\sfrac{1}{2}}}-\ket{P_{\sfrac{1}{2}}}$ dipole transition by
a single laser. So the fluorescence rate on this transition is
higher for the \ybion{172} isotope without hyperfine and Zeeman splitting compared to the \ybion{171}
ion where only a fraction of the $12$ possible transitions between $S_{\sfrac{1}{2}}$ and $P_{\sfrac{1}{2}}$
are driven in parallel. To profit from the better detection statistics we perform this part of the experiment using \ybion{172}.

To calibrate the measurements of the absent ion we perform the experiment with
$M_s$ transport operations $A\rightarrow B\rightarrow A$ skipped by intent (see
Fig.~\ref{fig:shuttleProof}).
As the ion is detected at position $A$ when it
should have been transported to position $B$, the detected fluorescence $F$ is
increasing with $M_s$.
The expected average photon count $F$ during an experiment of $M$ consecutive transfer
operations is given by $F(M_s) = M_s F_b + (M-M_s) F_d$,
with $F_b$ and $F_d$ being the average number of photons per detection flash collected
from a present (bright) and absent (dark) ion respectively.
If a fraction $f_s$ of transport operations $A\rightarrow B$ is failing, $f_s (M-M_s)$
additional bright and $f_s (M-M_s)$ fewer dark events are expected. For the
direction $B\rightarrow A$ one has to adjust $F$ by $f_s M_s F_d$ and $-f_s M_s F_b$ accordingly.
So the expected total photon number is
\begin{align}
F(M_s) =& F_b [M_s + f_s(M-2M_s)] +\notag\\ &F_d [(M-M_s) - f_s(M-2M_s)]\,.
\label{eq:transportFidelity}
\end{align}
In an additional run without an ion loaded to the trap we measured the average
photon number of an absent ion per flash to be $F_d=\num{23.5(3)}/M$.
From a fit of \eqref{eq:transportFidelity} to the data, we find the probability
for transporting the ion as intended per shuttling event to be
$1-f_s=0.998\left({}^{+2}_{-3}\right)$ --- corresponding to $8\left({}^{+12}_{-8}\right)$ out of $4000$ transport operations failing.

\section{Interpretation of the observed decoherence\label{appendix:decoherence}}
The measured internal state fidelity upon transport of $0.999994\left({}^{+6}_{-7}\right)$ is compatible with unity within one standard deviation.
Therefore, an unambiguous attribution of the Ramsey fringes'  loss of contrast to particular sources of decoherence is not possible.
Nevertheless, several processes can be ruled out as sources of decoherence from
the following discussion.

As the states of the hyperfine qubit are, for all practical purposes considered here, not subject to spontaneous emission, the
mechanism leading to decoherence is dephasing:
During the precession time the qubit state accumulates a phase $\Delta\varphi=\Delta\nu ~ t_p$
relative to the driving field of the $\pi/2$ pulses, where $\Delta\nu$ describes a
stochastically varying detuning of this field from the qubit transition.

Assuming a normal distribution
\begin{equation}
  \rho(\Delta\varphi) = \frac{1}{\sqrt{2\pi}\sigma_\varphi} e^{-\frac{(\Delta\varphi)^2}{2\sigma_\varphi^2}}
\end{equation}
of $\Delta\varphi$ over all measurements, a convolution of the Ramsey fringe $p_b(\varphi) = B + A \sin(\varphi - \Phi)
$ with $\Delta\varphi$ results in a contrast reduction by a factor of
$\exp(-\sigma_\varphi^2/2)$. The contrast
decay observed in this paper due to one qubit transport operation
yields $\sigma_\varphi=2\pi\times 6\left({}^{+8}_{-6}\right)\times10^{-4}$.

The magnetic field at the position
of the ion transport is $B = \SI{640}{\micro\tesla}$ and it's gradient in the
direction of the ion transport $dB/dx = \SI{10.6}{\milli\tesla\per\meter}$.
The magnetic field sensitivity of the qubit is $d\nu/dB = \SI{39.7}{\mega\Hz\per\tesla}$.
These values combined  yield a position sensitivity of the qubit
transition of $d\nu/dx = \SI{397}{\kilo\Hz\per\meter}$ in the direction of transport.

The repeated measurements of the Ramsey fringes with $2$ and $4000$ transport
operations are carried out interleaved, every single measurement being
synchronized to the phase of the power line. Therefore we exclude an additional
loss of contrast of the measurement using $4000$ transport operations due to
fluctuations of the magnetic field.

The electric currents caused by the transport potentials are much
too small to cause additional magnetic fields large enough to explain a dephasing
on the order of $10^{-4}$ per transport.
Furthermore, as the transport potentials are changed in a deterministic fashion,
the accompanying currents would result in a deterministic change of the magnetic
field that wouldn't cause dephasing. So even qualitatively a dephasing could only be caused by electronic
noise components, not by the potentials themselves.

Assuming a perfectly static magnetic field, a stochastic distribution of the
transport trajectories with a width of $\sigma_x$ would lead to dephasing due
to the magnetic gradient and the qubits residual magnetic field sensitivity.
With
\begin{equation}
  \sigma_x=\sigma_\varphi \left(2\pi t \frac{d \nu}{d x}\right)^{-1}
\end{equation}
and the duration of a single transport operation $t=\SI{12.8}{\micro\second}$,
the uncertainty in the transport trajectories required to produce the observed phase distribution would be
$\sigma_x=110\left({}^{+50}_{-110}\right)\mu\text{m}$. From our
observations of the ion after transport, we can exclude any value of $\sigma_x$
larger than a few $\mu\text{m}$. Therefore, possible stochastic variations of
the ion transport trajectory do not limit the measured internal state fidelity.
\begin{figure*}
  \includegraphics[]{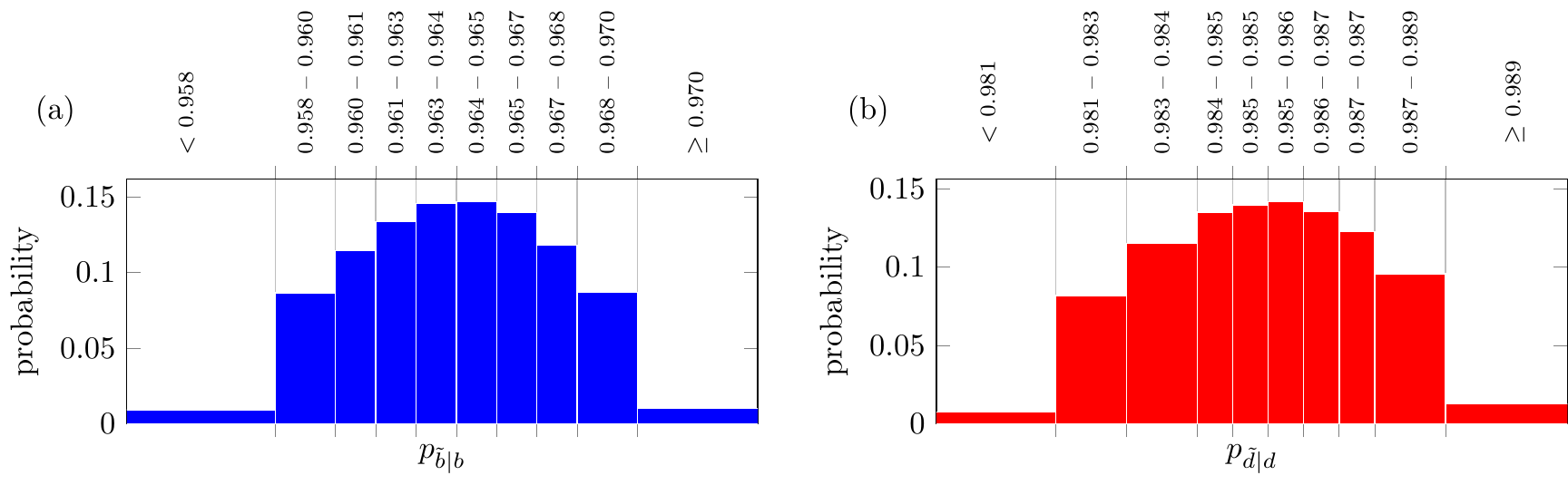}
\caption{State detection fidelities for $M=4000$ transport operations. The histograms
are generated by repeatedly selecting random sets from the calibration measurements.
The bin edges are selected such that the number of total events in the
single bins are closely matching while avoiding binning effects at the same time.
The first and last intervals are only shown in part.}
\label{fig:pXX}
\end{figure*}

\begin{figure}[b]
  \includegraphics[]{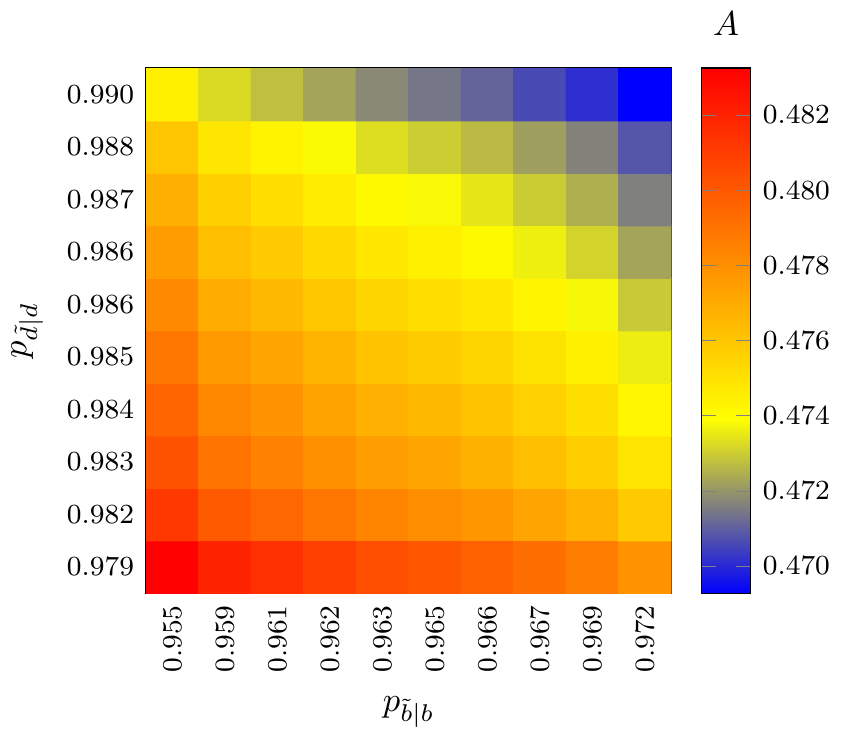}
\caption{The amplitude $A$ of the fitted Ramsey fringes depends on the assumed state
identification probabilities $p_{\tilde{b}|b}$ and $p_{\tilde{d}|d}$. If the states can be
identified with lower probability the likelihood analysis scales the measurement
data accordingly towards the extreme values $0$ and $1$ resulting in a higher
amplitude. To account for this effect the data analysis is carried out for 100
combinations of $p_{\tilde{b}|b}$ and $p_{\tilde{d}|d}$ with similar likelihoods.}
\label{fig:ramseySingleAmplitudes}
\end{figure}

None of the possible sources of dephasing discussed in this section is large
enough to explain the observed loss of internal state fidelity. We like to stress
that within the calculated error budget a finding of smaller transport induced
coherence loss is also not to be ruled out.

\section{Bootstrap analysis of the calibration data\label{appendix:bootstrap}}
The  state identification probabilities $p_{\tilde{b}|b}$ and $p_{\tilde{d}|d}$
are parameters of the data analysis. We estimate their uncertainties by applying
the bootstrapping resampling method~\cite{Efron1994} to the calibration data.
Figure~\ref{fig:pXX} shows the distributions of $p_{\tilde{b}|b}$ and $p_{\tilde{d}|d}$
obtained from $10000$ resampling runs. Binning of the data is done such that
every of the $10$ bins represents almost the same fraction $w(p_{\tilde{b}|b})$ ($w(p_{\tilde{d}|d})$) of results. So every combination
of $p_{\tilde{b}|b}$ and $p_{\tilde{d}|d}$ has almost the same significance of
$w(p_{\tilde{b}|b}, p_{\tilde{d}|d}) = w(p_{\tilde{b}|b})w(p_{\tilde{d}|d}) \approx 1/100$.

The maximum likelihood fit of the Ramsey fringes is carried out for every such combination
using the weighted mean for $p_{\tilde{b}|b}$ and $p_{\tilde{d}|d}$ in
the corresponding interval and the resulting profile likelihood function
$\mathcal{L}_M(A|p_{\tilde{b}|b}, p_{\tilde{d}|d})$ of the parameter $A$ is
calculated. Figure~\ref{fig:ramseySingleAmplitudes} shows the most likely
amplitudes for each combination of the state identification probabilities. If
one assumes smaller state identification probabilities the amplitude is
generally estimated to be higher.

\begin{figure}[b]
  \includegraphics[]{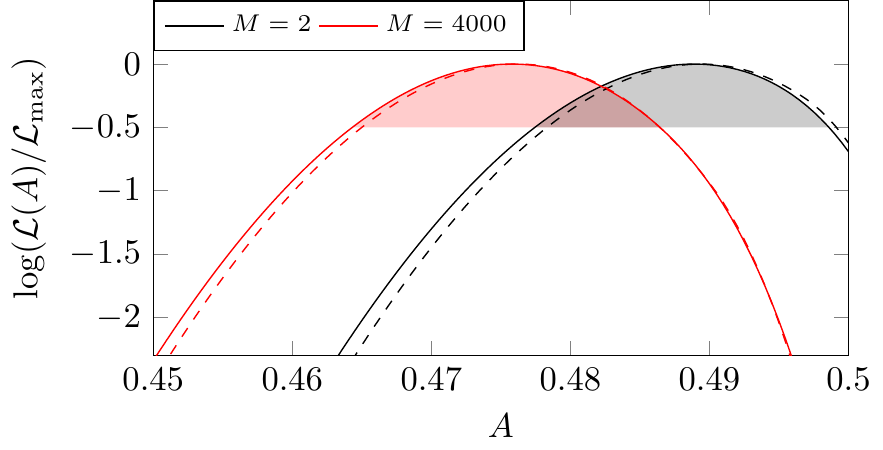}
\caption{Log likelihoods for the amplitude of the Ramsey fringes with $2$ and $4000$
transfer operations. The $68\%$ confidence intervals (shaded) are overlapping.
The log likelihoods are calculated by adding up the
likelihoods $\mathcal{L}_M(A|p_{\tilde{b}|b}, p_{\tilde{d}|d})$ obtained by data analysis for different
pairs of $p_{\tilde{b}|b}$ and $p_{\tilde{d}|d}$. For comparison the dashed curves
are showing the one pair of likelihoods $\mathcal{L}_M(A|p_{bb}, p_{dd})$ that is determined from the Ramsey fringes in main text, Fig.~2.}
\label{fig:ramseyFinalAmplitudes}
\end{figure}

Figure~\ref{fig:ramseyFinalAmplitudes} shows the log likelihood function $\mathcal{L}_M(A)$
obtained by summing up the single likelihoods
\begin{equation}
\mathcal{L}_M(A) = \sum_{p_{\tilde{b}|b}, p_{\tilde{d}|d}} w(p_{\tilde{b}|b}, p_{\tilde{d}|d}) \mathcal{L}_M(A|p_{\tilde{b}|b}, p_{\tilde{d}|d})
\end{equation}
Again one can see that the most likely amplitude of the Ramsey fringes for more transport
operations is slightly smaller. The $1 \sigma$ confidence interval of the
amplitude for $M=2$ and $M=4000$ are overlapping. By comparing the likelihoods obtained with and without usage of the bootstrapped
distributions $p_{\tilde{b}|b}$ and $p_{\tilde{d}|d}$ one can see that the inclusion of this statistical error source
is broadening and slightly shifting the maximum of
the calculated likelihoods.

\section{Preparation errors\label{appendix:preparationErrors}}
The identification probabilities $p_{\tilde{b}|b}$ and $p_{\tilde{d}|d}$ are
determined under the assumption that the bright and dark states can be
prepared with unit fidelity.

If the states $\ket{0}$ and $\ket{1}$ are not correctly prepared, the identification
probabilities are systematically biased towards lower values. The usage of low biased values
in the data analysis will result in systematically overestimated fringe contrasts
(compare Fig.~\ref{fig:ramseySingleAmplitudes}). If the state preparation ($\fidelity[p]$) and $\pi$ pulse ($\fidelity[\pi]$) fidelities are known, unbiased values $\overline{p}_{\tilde{b}|b}$
and $\overline{p}_{\tilde{d}|d}$ can be calculated as
\begin{align}
   \overline{p}_{\tilde{b}|b}  &= \frac{p_{\tilde{d}|d}(2 \fidelity[p] \fidelity[\pi] - \fidelity[p] -\fidelity[\pi]) + \fidelity[p] (1 - 2\fidelity[\pi]-p_{\tilde{b}|b}) + \fidelity[\pi]}{(1-2 \fidelity[p]) \fidelity[\pi]}\label{eq:pBB_bias}\\
   \overline{p}_{\tilde{d}|d}  &= \frac{p_{\tilde{d}|d}(-2 \fidelity[p] \fidelity[\pi] + \fidelity[p] + \fidelity[\pi] - 1) + (\fidelity[p]-1)(p_{\tilde{b}|b}-1)}{(1-2 \fidelity[p]) \fidelity[\pi]}\label{eq:pDD_bias}
\end{align}
The statistics approach used in this paper includes the effect of the imperfect
identification probabilities.
By connecting identified states to projected states,
the probability densities for a
given measurement result are effectively rescaled.

An alternative approach to take the limited state identification
probabilities into consideration is to first apply a
linear transformation
\begin{equation}
  \underbrace{
  \left(\begin{array}{c}
   b\\
   d
 \end{array}\right)
 }_{\vec{r}}
 =
 {
\underbrace{
 \left(\begin{array}{cc}
  p_{\tilde{b}|b} & 1-p_{\tilde{d}|d} \\
  1-p_{\tilde{b}|b} & p_{\tilde{d}|d} \\
 \end{array}
 \right)%
 }_{ M(p_{\tilde{b}|b}, p_{\tilde{d}|d})}
 }^{-1}
 \underbrace{
 \left(\begin{array}{c}
   \tilde{b}\\
   \tilde{d}
 \end{array}\right)
 }_{\vec{\tilde{r}}}
 \label{eq:linear_tagged_state}
\end{equation}
to the identified measurement results
$\vec{\tilde{r}} = (\tilde{b}, \tilde{d})^T$ and
then employ the beta probability function on the rescaled measurement results $b$ and $d$. The transformation matrix $M(p_{\tilde{b}|b}, p_{\tilde{d}|d})$ is determined
by the identification probabilities. Even though this approach distorts
the probability density $\rho(p_b|\tilde{b},N)$ and systematically underestimates statistical errors by not
taking into account the broadening of the distribution due to the
stochastic interpretation of projected states, described by binomial
distributions with probabilities $p_{\tilde{b}|b}$ and $p_{\tilde{d}|d}$,
the
most likely value for the probability density of a measured state remains unbiased.

Using this simplified method we discuss the effects of biased detection
probabilities on the bias for determination of the internal state fidelity
$\fidelity$:

If $\overline{A}$ and $\overline{B}$ are the
amplitude and offset of the Ramsey fringe,
an outcome
\begin{equation}
  \vec{\overline{r}}_\text{max}
  = N
  \left(\begin{array}{c}
    \overline{B}+\overline{A}\\
    \overline{B}-\overline{A}
  \end{array}\right)
\end{equation}
at the fringe maximum
for $N$ measurements is to be expected,
a reading of
$\vec{\overline{r}}_\text{min}=N(\overline{B}-\overline{A}, \overline{B}+\overline{A})^T$
at the minimum respectively. Due to the limited state identification
probabilities these measurement outcomes will be identified as
$\vec{\tilde{r}}_\text{max} = M(\overline{p}_{\tilde{b}|b}, \overline{p}_{\tilde{d}|d}) \vec{\overline{r}}_\text{max}$
and
$\vec{\tilde{r}}_\text{min} = M(\overline{p}_{\tilde{b}|b}, \overline{p}_{\tilde{d}|d}) \vec{\overline{r}}_\text{min}$. A transformation according to \eqref{eq:linear_tagged_state} based
on biased state identification probabilities yields
$\vec{r}_\text{min/max} = [M({p}_{\tilde{b}|b}, {p}_{\tilde{d}|d})]^{-1} M(\overline{p}_{\tilde{b}|b}, \overline{p}_{\tilde{d}|d}) \vec{\overline{r}}_\text{min/max}$.
Using \eqref{eq:pBB_bias} and \eqref{eq:pDD_bias} one finds, that the factor between
unbiased ($\vec{\overline{r}}$) and biased ($\vec{r}$) values
depends on the state preparation and $\pi$ pulse fidelities only and does not depend
on $p_{\tilde{b}|b}$ or $p_{\tilde{d}|d}$. The biased
amplitude $A$ of the Ramsey fringes
is given by half the difference of bright events in the maximum
and minimum of
the Ramsey fringe $\frac{1}{2}(\vec{r}_\text{max}-\vec{r}_\text{min})_1$:
\begin{align}
A &= \frac{1}{2}\left\{[M({p}_{\tilde{b}|b}, {p}_{\tilde{d}|d})]^{-1} M(\overline{p}_{\tilde{b}|b}, \overline{p}_{\tilde{d}|d}) \frac{\vec{\overline{r}}_\text{max}-\vec{\overline{r}}_\text{min}}{N}\right\}_1\nonumber\\
&= \frac{\overline{A}}{\fidelity[\pi](2\fidelity[p]-1)}\label{eq:fidelity_independent}
\end{align}
The amplitude is biased towards higher values, but for the determination of the
internal state fidelity during ion transport
 $\fidelity=\sqrt[4000-2]{A_2/A_{4000}}$ only the ratio of the amplitudes for $M=2$
and $M=4000$ is of interest and the factor $[\fidelity[\pi](2\fidelity[p]-1)]^{-1}$
in \eqref{eq:fidelity_independent} cancels.

As noted before the linear transformation used in the reasoning above distorts the
probability density function and an exact statistical treatment of the problem reintroduces a small dependence
of the fidelity $\fidelity$ on the state identification probabilities. We assume
a preparation fidelity of $\fidelity[p] >1-\num{1e-4}$ and a fidelity of the
fault tolerant BB1RWR $\pi$ pulse of $\fidelity[\pi] > 1-\num{1.3e-4}$.
The corresponding maximal bias of the identification probabilities are
$\num{-9.5e-5}$ ($p_{\tilde{b}|b}$) and $\num{-2.2e-4}$ ($p_{\tilde{d}|d}$). A
data analysis assuming these biases yields numerically insignificant
changes ($-\num{2.4e-8}$) to the overall result for the internal state fidelity during ion transport.
This still holds when infidelities of $\fidelity[\pi]$ and $\fidelity[p]$ of $\num{1e-3}$ are
assumed ($-\num{2.2e-7}$).

\end{document}